# Easy-to-configure zero-dimensional valley-chiral modes in a graphene point junction

Konstantin Davydov[1], Xi Zhang[1], Wei Ren[1], Matthew Coles[1], Logan Kline[1], Bryan Zucker[2], Kenji Watanabe[3], Takashi Taniguchi[4], Ke Wang[1*]

[1]School of Physics and Astronomy, University of Minnesota, Minneapolis, Minnesota 55455, USA

[2]Department of Physics, The Ohio State University, Columbus, Ohio 43210, USA

[3]Research Center for Electronic and Optical Materials, National Institute for Materials Science; 1-1 Namiki, Tsukuba, 305-0044, Japan.

[4]Research Center for Materials Nanoarchitectonics, National Institute for Materials Science; 1-1 Namiki, Tsukuba 305-0044, Japan.

**The valley degree of freedom in 2D materials can be manipulated for low-dissipation quantum electronics called valleytronics. At the boundary between two regions of bilayer graphene with different atomic or electrostatic configuration, valley-polarized current has been realized. However, the demanding fabrication and operation requirements limit device reproducibility and scalability toward more advanced valleytronics circuits. We demonstrate a new device architecture of a point junction where a valley-chiral 0D PN junction is easily configured, switchable, and capable of carrying valley current with an estimated polarization of ~80%. This work provides a new building block in manipulating valley quantum numbers and scalable valleytronics.**

The band edges of typical 2D semimetals [1] and semiconductors are found at two corners of the first Brillouin zone commonly referred to as K and K′ valleys [2,3]. In monolayer transition metal dichalcogenides, dynamic valley polarization can be achieved with optical excitations [4–7]. At a 1D boundary of two bilayer graphene (BLG) domains (natural or gate-defined), topological valley-chiral states can be realized [8–14] when two critical conditions are simultaneously met: (A) carrier density reaches zero in both the two domains and the boundary, and (B) the Berry phase [15,16] of each valley switches sign at the boundary. The valley-chiral 1D state has been experimentally demonstrated in two major device architectures: gate-defined [8–10,13] or naturally occurring [11,12,14,17,18]. The former scheme utilizes two pairs of split gates perfectly aligned on top of each other, requiring extreme precision in angle and overlay alignment for device fabrication. The charge neutrality in the 1D channel must also be ensured by tuning an additional global gate. The latter scheme relies on natural domains of opposite Bernal-stacking orders. While the device fabrication no longer requires precise overlay alignment, the existence of a natural domain boundary is rare and with its pattern naturally pre-defined [11,12,14,20,21], requiring optical identification and limits versatile circuit design. These successful experiments provide important proof-of-the-principle demonstration for valleytronics [8–14,19–29], but scalability toward the valley-chiral circuit network desires a new device scheme with higher yield and simpler fabrication and operation.

In this work, we have developed a novel device architecture of a point junction (PJ), where a PN junction [30] and a quantum point contact [27,31–35] (QPC) are simultaneously defined at the center of the device (Fig. 1a). The PJ can be electrostatically configured easily without the need for precise overlay alignment nor precise electrostatic tuning. The hexagonal boron nitride (hBN)-encapsulated [36] BLG stacks are transferred on top of a pair of pre-deposited bottom gates, each with a 1-2 µm width and a ~100 nm gap in between them (Fig. 1b to d). The top surface of the bottom gates has been characterized to be atomically flat (Fig. 1b, c) to minimize lattice strain or a remote defect that may cause inter-valley scattering in the graphene above. A pair of top gates is subsequently deposited with similar geometry but orthogonal orientation to the bottom gates (Fig. 1d). A pair of 1D edge contacts is then fabricated [37] at each of the four corners of the devices, and reactive ion-etching defines the final geometry of the device (Fig. 1a).

We first demonstrate how a point junction can be electrostatically defined. An arbitrary combination of gate voltages can be applied to the top and bottom gates which we hereby denote as ($V_{t1}$, $V_{t2}$, $V_{b1}$, $V_{b2}$). This electrostatically divides the device into a 2×2 grid with each region's carrier density and gap size [38–40] determined by a given pair of the four parameters. A point junction is easily configured by setting $V_{t1} = -V_{b1}$, and $V_{t2} = -V_{b2}$. As an example (Fig. 1e), at ($V_{t1}$, $V_{t2}$, $V_{b1}$, $V_{b2}$) = (-2.5V, -5V, +2.5V, +5V), the top left (bottom right) region of the device is N (P) doped while the other two regions are gapped (with the same displacement field direction). At the center of the device and the common corner of all four regions, a PN junction (along the direction across P and N doped regions) and a constriction (quantum point contact, along the direction across two insulating regions) are simultaneously defined - a single point in space where P and N region meets to define a zero-dimensional PN junction. This is a natural consequence of the electrostatic potential from the device architecture, without requiring elaborate device tuning or precise overlay alignment.

The quantitative details of each region can be seen with a band-diagram along two diagonal directions of the device. Between far corners of the P and N regions, the gap size stays the same while the carrier type changes at the PJ. The gap size of the two depleted regions changes at the PJ, but it does not close (the displacement field does not switch sign). We therefore do not expect valley-chiral states in this "trivial" configuration, i.e. the electrons from both valleys should contribute to the transport equally (Fig. 1f, g) via non-chiral quantum tunneling. We perform a 4-probe measurement of quantum transport in this

"trivial" configuration both as a control study to show whether valley-chiral states (when configured) will make a qualitative difference and as a characterization of realistic quantum tunneling across the PJ.

To do this, we set the top gates' voltage to be $(V_{t1}, V_{t2}) = (-2.5V, -5V)$, while measuring the 4-probe resistance (Fig. 1h) from the P to the N side of the device as a 2D sweep of $(V_{b1}, V_{b2})$. The pre-set top gate voltages set the target doping and gap size for each region of the eventual PJ (for tunability of each device region). The 2D sweep (instead of requiring three or more parameters) allows us to accurately configure and identify the PJ when measured resistance reaches its maximum value as a consequence of two insulating regions being established simultaneously. The two parameters $(\Delta V_{b1}, \Delta V_{b2})$ denote the difference between actual voltage applied and the expected value of $(V_{b1}, V_{b2}) = (+2.5V, +5V)$ at which the PJ is defined with $(\Delta V_{b1}, \Delta V_{b2}) = (0, 0)$ corresponding to the formation of two insulating regions and successful establishment of the PJ. A resistance peak is observed at $(\Delta V_{b1}, \Delta V_{b2}) = (0, 0)$, as expected from the suppression of current via the two regions that become insulating. This resistance characterizes the quantum tunneling process across the PJ, and monotonically increases with the displacement field (Fig. 1i) as the tunnel barrier at the PJ increases. The resistance's dependence on DC voltage (Fig. 1j) shows non-Ohmic behavior, and begins to decrease beyond a 5 mV – 10 mV bias. This is consistent with the estimated range of gap sizes of $\Delta = 10$ mV to 20 mV and previously reported values [39–41] at displacement fields, $D/\varepsilon_0$, ranging from 0.14 to 0.26 V/nm.

With the above characterization of trivial tunneling across the PJ, we now add the new component of the valley-chiral modes. This can be done by applying a different gate configuration. An example is shown for $(V_{t1}, V_{t2}, V_{b1}, V_{b2}) = (+2.5V, -5V, -2.5V, +5V)$. In contrast to the previous configuration, the carrier density and displacement field (and gap) switch sign simultaneously at the PJ (Fig. 2a) satisfying both requirements for valley-chiral states. Similarly, this is automatically ensured by device electrostatics instead of precise tuning and overlay alignment.

If, for example, a current is driven from P to N. While carriers in both K and K' valleys (Fig. 2c) can contribute via realistic quantum tunneling (as characterized in Fig. 1), only carriers from the K valley (Fig. 2b) can pass through the 0D valley-chiral modes. Similar to the previous measurement, we pre-set the top gates' voltage to be $(V_{t1}, V_{t2}) = (+2.5V, -5V)$ while measuring the 4-probe resistance (Fig. 2d) from the P to the N side of the device as a 2D sweep of $(V_{b1}, V_{b2})$. A resistance peak is observed near $(\Delta V_{b1}, \Delta V_{b2}) = (0, 0)$ with respect to the expected $(V_{b1}, V_{b2}) = (-2.5V, +5V)$. The resistance itself (Fig. 2e) is significantly smaller than that of the "trivial" configuration (Fig. 1i) with the same gap sizes (data points with matching color) for the insulating regions. Its nearly zero-dependence on DC bias (Fig. 2f) suggests ohmic behavior that is expected from a ballistic 1D channel. These observations qualitatively confirm the existence of the valley-chiral modes, naturally defined by the PJ electrostatics, which can be easily fabricated, configured, and tuned.

The measured conductance in this configuration is contributed by two parallel channels: (1) realistic quantum tunneling across the PJ that applies to both valleys via which the current is non-chiral, and (2) valley-chiral states with a ballistic conductance of $4e^2/h$ and are exclusive to K valley via which the current is valley-chiral. The latter promises 100% valley polarization and $4e^2/h$ quantized conductance while the co-existence of the former results in reduced valley polarization and a measured total PJ conductance higher than $4e^2/h$. As such, we model the two co-existing mechanisms as conducting channels in parallel from which we estimate a valley polarization of ~ 80% via formula (see SI: section S3) *VP = (1 + R/R$_0$)/2* (Fig. 2g). Despite this limitation on a single PJ, valley polarization can be further improved by operating multiple valley-chiral PJs in series with additional top and bottom gates that are easily scalable.

The non-chiral quantum tunneling that compromises the valley-polarization can be further distinguished into two types. Type-1: tunnel current via the region directly under/above the local gates away from the PJ where the size of bandgap is constant. This leakage channel can be suppressed by further increasing the bandgap with a larger displacement field. Type-2: tunnel current via the region in between the gate separations in the vicinity of the PJ where the size of bandgap is determined by the fringing field, and has strong spatial dependence that is different in the "trivial" and "non-trivial" configurations (see SI: section S4). In the "trivial" configuration, Type-2 tunneling is weaker with a finite tunnel barrier near the PJ, and thus can be further suppressed with a higher $D$ field which is consistent with the comparable rapid monotonic dependence shown in Fig. 1i. In the "non-trivial" configuration, the tunneling is impossible to eliminate with electrostatics alone as the effective tunnel barrier diminishes near the vicinity of the PJ. As a result, the estimated *VP* increases slowly with the displacement field (Fig. 2g) as Type-1 tunneling is being suppressed effectively while Type-2 is not.

To estimate the Type-2 tunneling current, we subtract the conductance of the PJ between when it is configured with and without the valley-chiral state. The size of the gaps as well as the Type-1 tunneling currents are the same in the two configurations. The conductance after subtraction (Fig. 2h) is observed to be 10% - 20% higher than $4e^2/h$ (a value expected if there was no leakage current due to tunneling which characterizes the Type-2 tunneling current near the vicinity of the PJ in the valley-chiral configuration which is a primary limiting factor for valley polarization).

While such Type-2 tunneling in a valley-chiral PJ cannot be eliminated by electrostatics due to band-gap closing near the PJ, it can be suppressed by a Landau gap at zero charge density ($v = 0$) with an out-of-plane magnetic field $B$ applied [42]. At low field, the measured resistance of the PJ with and without valley-chiral modes increases with the $B$-field as the zero-energy Landau gaps help to suppress both Type-1 and Type-2 tunneling. At high field, the resistance of the valley-chiral PJ starts to demonstrate Shubnikov-de Haas (SdH) oscillations (Fig. 3a) with the oscillation minima close to $h/(4e^2)$ (dashed lines). Each color plot is taken at different displacement fields offset with equal separation.

The observation is consistent with the expectation from valley-specific Chern numbers in bilayer graphene. At $B = 0$ and $D > 0$, while the net Chern number of BLG as a whole is 0 (making it topologically trivial), the K (K′) valley each has an effective Chern number (valley Chern number) of 1 (-1) that leads to valley-chiral 1D modes at the point junction. Similarly, at finite magnetic field, while BLG as a whole has its Chern number equal to Landau level indices, $N$, the K (K′) valleys each have a net valley Chern number of $N + 1$ ($N - 1$). The valley-specific net Chern number leads to valley-specific QH edge states.

To elaborate on this, we illustrate separately the spatial distribution of compressible states (CS, red: P type, blue: N type, yellow: valley-chiral) and incompressible states (white) for the K valley (Fig. 3b, d) and the K′ Valley (Fig. 3c, e). At high magnetic field, exactly two QH edge states exist at the point junction with opposite current direction for charges in different valleys (valley-chiral). The existence of these valley-chiral QH states is consistent with a series of resistance dips of $h/(4e^2)$ observed when the bulk of both the P and N regions are in a gap and are incompressible (Fig. 3b, c). The two valley-chiral states ballistically connect the P and the N side of the device independent of the bulk Landau level indices. When the bulk P and N regions become compressible (Fig. 3d, e), a resistance peak is measured due to back-scattering in the bulk region.

As a control comparison, we perform the same measurement with the "trivial" configuration (Fig. 3f) where the two insulating regions and the two doped regions are defined with a $D$ field of the same sign. The QH states' distributions are the same for both valleys (fig. 3g, h) because the valley Chern number does not change across any boundary across adjacent device regions, and the valley-chiral QH states are not expected (Fig. 3g, h). As the $v = 0$ Landau gap increases, the tunneling between QH states [43,44] on the P and N sides becomes more suppressed leading to a measured resistance that sharply increases with $B$ field instead of the oscillatory behavior.

In summary, we have developed a novel device architecture of a graphene point junction. We demonstrate for the first time, the transport signature of: (1) An electrostatically defined zero-dimensional

PN junction. (2) A zero-dimensional valley-chiral valve with estimated valley-polarization of 80%. (3) Topologically protected valley-chiral quantum Hall states with ~100% polarization under high magnetic field. All above states can be easily configured without demanding device fabrication protocol and precise electrostatic tuning. The ease of configuration and reproducibility allows valley-chiral valves to be configured in series to further improve the valley polarization, or in parallel for valley logic device. The device provides a new platform to study valley-polarized quantum phenomena in graphene and a robust new device component toward scalable valleytronics.

We thank Paul Crowell, Alex Kamenev, Tony Low and Duarte Pereira de Sousa for helpful discussions. K.D., X.Z., W.R., M.C., L.K., B.Z., K.W. are supported by the National Science Foundation CAREER Award NSF-1944498. Portions of this work were conducted in the Minnesota Nano Center, which is supported by the National Science Foundation through the National Nanotechnology Coordinated Infrastructure (NNCI) under Award Number ECCS-2025124. K.W. and T.T. acknowledge support from the JSPS KAKENHI (Grant Numbers 20H00354, 21H05233 and 23H02052) and World Premier International Research Center Initiative (WPI), MEXT, Japan."

*kewang@umn.edu

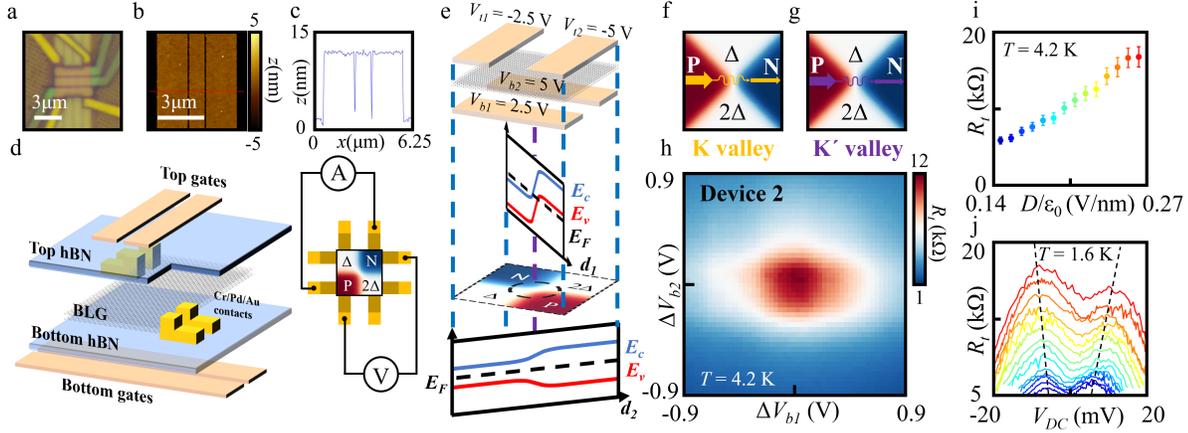

**Figure 1. Electrostatically defined zero-dimensional PN junction**. (a) Optical image of a PJ device. (b) AFM topography of typical pre-patterned bottom gates with (c) 1D cut (along the red line in (b)) demonstrating the atomically flat gate surface critical for ballistic electrostatic manipulation. (d) Schematics of the device architecture. A pair of metal top gates and a pair of bottom gates define two voltages. (e) Electrostatics of a "trivial" PJ configuration without the valley-chiral modes. As an example, gate voltages ($V_{t1}$, $V_{t2}$, $V_{b1}$, $V_{b2}$) = (-2.5V, -5V, +2.5V, +5V) electrostatically divide the BLG into two N-(P-) doped and two insulating (with $\Delta$ and $2\Delta$ bandgaps) regions with 1D band diagrams along two diagonal directions and the 2D carrier density profile shown below. The bandgap does not close at the PJ in the center of the device. (f and g) Carriers from the K and K′ valleys contribute equally to the measured tunneling current across the PJ without valley chirality. (h) Qualitative feature of measured 4-probe resistance of a typical PJ as voltages on the two bottom gates are tuned away from ($V_{b1}$ = +2.5V, $V_{b2}$ = +5V) by $\Delta V_{b1}$ and $\Delta V_{b2}$. The resistance peak is observed when a PJ is established at $\Delta V_{b1} = \Delta V_{b2} = 0$, and decreases away from the peak as either of the two insulating regions becomes doped. (i) Measured 4-probe PJ resistance as a function of the displacement field, and (j) the PJ's resistance as a function DC bias, $V_{DC}$, measured at different $D$ fields from $D/\varepsilon_0 = 0.14$ V/nm (blue) to $D/\varepsilon_0 = 0.26$ V/nm (red) with an increment of 0.01 V/nm. The PJ resistance increases with a $D$ field applied as tunneling is suppressed with a higher barrier. Its bias dependence is non-ohmic and starts to decrease at a bias voltage comparable to the BLG gap size under such a $D$ field.

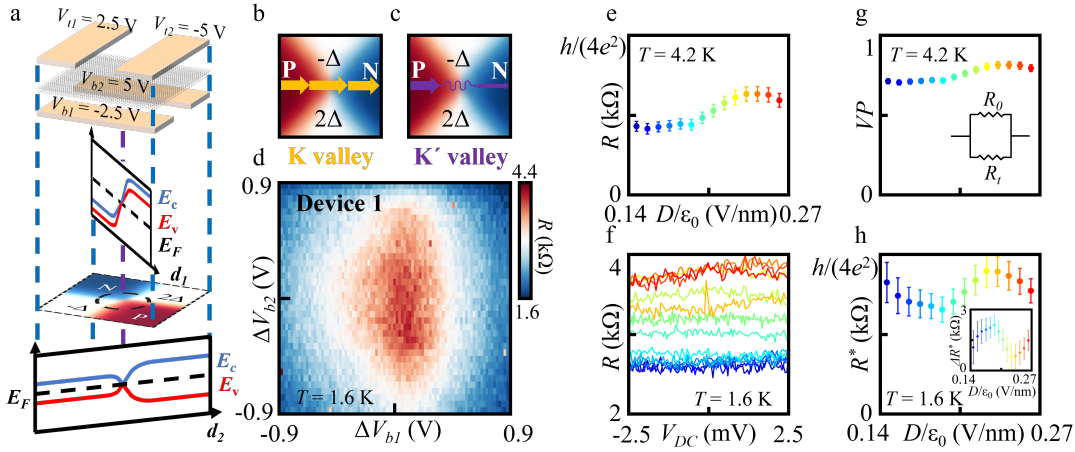

**Figure 2. Easy-to-configure zero-dimensional valley valve.** (a) Electrostatics of the "non-trivial" PJ configuration with the valley-chiral modes. As an example, gate voltages ($V_{t1}$, $V_{t2}$, $V_{b1}$, $V_{b2}$) = (+2.5V, -5V, -2.5V, +5V) electrostatically divide the BLG into two N-(P-) doped and two insulating (with -$\Delta$ and 2$\Delta$ bandgaps) regions with 1D band diagrams along two diagonal directions and the 2D carrier density profile shown below. The bandgap and carrier density flip signs simultaneously at the center of the device where the PJ is defined simultaneously satisfying both conditions for valley-chiral modes. (b), (c) While realistic quantum tunneling is expected from both valleys, current from the P to the N region is primarily carried by valley-chiral 1D modes with carriers exclusively from the K valley. (d) Qualitative feature of measured 4-probe resistance of a typical PJ, as voltages on the two bottom gates are tuned away from ($V_{b1}$ = -2.5V, $V_{b2}$ = +5V) by $\Delta V_{b1}$ and $\Delta V_{b2}$. The resistance peak is observed when a valley-chiral PJ is established at $\Delta V_{b1} = \Delta V_{b2} = 0$ and decreases away from the peak as either of the two insulating regions become doped. (e) Measured 4-probe PJ resistance as a function of the displacement field, $D$, (f) and as a function DC bias, $V_{DC}$. The measured PJ resistance shows ohmic behavior as expected from ballistic chiral channel and has weaker dependence on the $D$ field as the effective barrier for Type-2 tunneling vanishes near the PJ. (g) Estimated valley polarization and (h) net PJ resistance characterized at different $D$ fields from $D/\varepsilon_0$ = 0.14 V/nm (blue) to $D/\varepsilon_0$ = 0.26 V/nm (red) with an increment of 0.01 V/nm. The Type-2 tunneling in the "non-trivial" mode can be estimated by the difference (h inset) between $h/(4e^2)$ and the net PJ resistance.

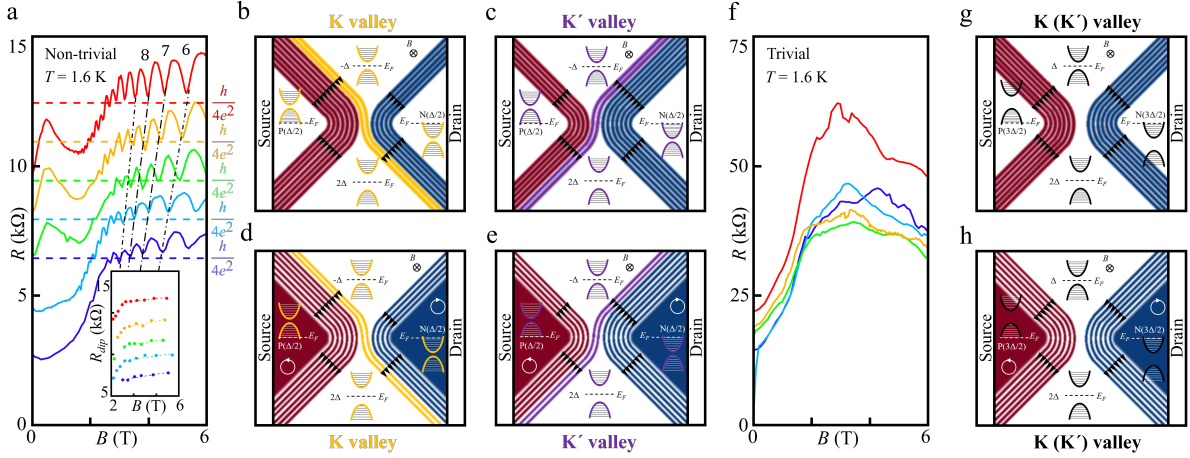

**Figure 3. Valley-chiral quantum Hall states with topologically protected high polarization.** (a) Shubnikov-de Haas oscillations in the resistance of the valley-chiral PJ at different displacement fields from 0.19 V/nm (purple) to 0.23 V/nm (red) with a 0.01V/nm increment. The plots are offset vertically by 1.5 kΩ and the colored dashed lines refer to $h/(4e^2)$ – the expected resistance from valley-chiral states. The dip positions (black dash-dotted) are consistent with different bulk doping at different $D$ fields. Inset: The oscillation dips' resistances show a weak $B$ field dependence with values around $h/(4e^2)$. (b to e) The spatial distribution of compressible states (red: P type; blue, N type; yellow, zero-energy K valley-chiral; purple: zero-energy K′ valley-chiral) and incompressible states (white). At integer bulk filling, electronic transport occurs exclusively via valley-chiral states both for the K valley (b) and the K′ valley (c), leading to an observed dip resistance of $h/(4e^2)$. At partial bulk filling, backscattering resistance in the bulk region in series to valley-chiral resistance $h/(4e^2)$ gives rise to the observation of resistance peaks. (f) Control experiment performed with all four regions defined with a $D$ field of the same sign ("trivial" configuration) where the valley-chiral QH states are expected for neither of the two valleys (g and h). As the ν = 0 Landau gap increases with larger $B$ fields, the measured resistance grows sharply and then saturates in contrast to the oscillatory behavior.

# Easy-to-configure zero-dimensional valley-chiral modes in a graphene point junction

Konstantin Davydov[1], Xi Zhang[1], Wei Ren[1], Matthew Coles[1], Logan Kline[1], Bryan Zucker[2], Kenji Watanabe[3], Takashi Taniguchi[4], Ke Wang[1*]

[1]School of Physics and Astronomy, University of Minnesota, Minneapolis, Minnesota 55455, USA

[2]Department of Physics, The Ohio State University, Columbus, Ohio 43210, USA

[3]Research Center for Electronic and Optical Materials, National Institute for Materials Science; 1-1 Namiki, Tsukuba, 305-0044, Japan.

[4]Research Center for Materials Nanoarchitectonics, National Institute for Materials Science; 1-1 Namiki, Tsukuba 305-0044, Japan.

**S1. Sample Preparation, Device Fabrication, and Measurement Methods**

To fabricate the devices, two parallel pieces of 1-2 µm wide metal gates (serving as $V_{b1}$, $V_{b2}$) with a separation of ~100 nm and consisting of Cr/Pd-Au alloy (1 nm/7 nm) are deposited on a 285 nm thick $SiO_2$ substrate (Fig. 1b, c). The Pd-Au alloy (40% Pd / 60% Au) is chosen to reduce surface roughness compared to the conventional Au deposition. Then, the gates are annealed in a high-vacuum environment at 350 °C for 15 minutes to remove surface residue and to ensure an atomically flat topography. After the annealing, the gates are examined (Fig. 1b, c) under an atomic force microscope (AFM) confirming their flatness and their lack of contamination.

Top hBN, bilayer graphene (BLG), and bottom hBN flakes (Fig. 1d) were consecutively picked up using polypropylene carbonate (PPC) and polydimethylsiloxane (PDMS) stamps before transferring onto the pre-deposited gates following the standard dry transfer technique [1,2]. The sample was rinsed in acetone and isopropanol to remove the PPC residue and then annealed in a high-vacuum environment at 350 °C for 15 minutes. AFM topographies of the sample were taken to ensure that it was free of air bubbles and local atomic strain. After that, a pair of parallel metal top gates (serving as $V_{t1}$, $V_{t2}$) separated by a ~100 nm gap and perpendicular to the bottom gates were fabricated by electron-beam lithography and Cr/Pd/Au (1 nm/5 nm/14 nm) deposition. Following that, electrical contacts to gates and ohmic contacts [3] to 1D boundaries of the bilayer graphene were made by electron-beam lithography, dry-etching, and subsequent metal deposition (Cr/Pd/Au, 1 nm / 5 nm / >120 nm). Lastly, the final lateral geometry (Fig. 1a) of the device was defined by the ultimate round of dry etching.

The electrical- and magneto-transport measurements for multiple devices studied in this work are performed in a Cryostation® s50 (Montana Instruments) at the base temperature of 4.2 K and in a TeslatronPT (Oxford Instruments) at the base temperature of 1.6 K with magnetic fields up to 6 T. The measurements were carried out with AC excitations of 17.777 Hz. The voltage drops and currents were measured using SR830 lock-in amplifiers (Stanford Research Systems) and current preamplifiers (DL: Model 1211). The DC bias voltage was controlled by a waveform generator (Keysight: Model 33500B). The gate voltage was applied from various voltage sources (Keithley Instruments: Model 2400, Yokogawa: Model GS200, QDevil: Model Q302 QDAC, and a custom-made DAC). Depending on the specific device architecture (number of parallel gates that can be utilized), the resistance (four-probe for every device except Device 4) of one PJ and/or two PJs in series was measured. In the latter case, we expect the 0D chiral modes to facilitate a

$4e^2/h$ conductance across the device (as expected from ballistic quantum conductance) while we expect the non-chiral trivial quantum tunneling resistance, $R_t$, to be a sum from the tunneling resistance of two PJs (effectively as two resistors in series). In this case, $R_t$ of a single PJ can be estimated as one-half of the resistance of the two PJs in series.

### S2. Characterization of Type-1 and Type-2 Non-chiral Tunneling Conductance at Zero $B$ Field

The non-chiral quantum tunneling can be attributed to two types. Type-1: tunnel current via insulating regions directly under/above local gates, away from the PJ, where the size of the bandgap is constant. Type-2: tunnel current via the region in between the gate separation in the vicinity of the PJ. In the "trivial" configuration, the band gap is non-zero in all four regions of the device. As a result, there will be a finite tunnel barrier (Fig. S1a) between the P and N regions with a height proportional to the bandgap size. Applying a $D$ field, thus increasing the bandgap, suppresses both Type-1 and Type-2 tunneling which results in a monotonic growth of $R_t$ as a function of $D$ (Fig. 1i).

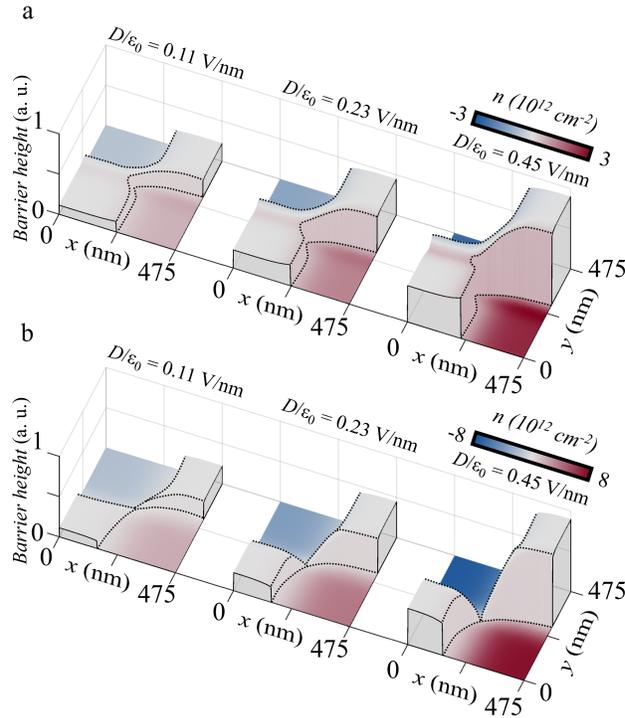

**Figure S1. COMSOL Electrostatics Simulation of Potential Barrier Height in "Trivial" and Valley-Chiral Configurations of PJ.**

(a) Potential barrier height (in arbitrary units) and the charge carrier density ($n$) near the PJ in the "trivial" case at different displacement fields (the values correspond to the displacement field in the Δ region). (b) Same as (a), but for the valley-chiral PJ. Unlike (a), the P and N regions are no longer separated by a finite barrier. The magnitude of the displacement fields in the bulk insulating regions is set to be the same for both configurations. The dashed lines highlight the shape of the tunnel barriers.

In contrast, in the valley-chiral configuration, the bandgap always closes at the PJ independent of the $D$ field applied. As a result, only Type-1 tunneling can be suppressed (Fig. S1B). The measured resistance of the valley-chiral PJ therefore exhibits a much weaker dependence on $D$ that eventually saturates (a higher $D$ field no longer helps).

At $B = 0$ and finite $D$ field (main manuscript: Fig. 2), the measured conductance ($1/R$) across the PJ with the valley-chiral configuration is composed of two parallel conducting channels (Fig. S2): valley-chiral ballistic 1D states ($1/R_0$), Type-1 + Type-2 ($1/R_t$) non-chiral tunneling, respectively. The upper bound of $R_t$ is directly characterized by the measured resistance of the PJ in the "trivial" configuration where the 0D chiral states ($1/R_0$) are absent due to the gap never closing in the PJ or across the entire device. The main manuscript shows the plot (Fig. 1h) of the net resistance of the valley-chiral PJ $1/R^* = 1/R - 1/R_t = 1/R_0 + \Delta(1/R_t)$ where $\Delta(1/R_t)$ is the increase in tunneling compared to the "trivial" PJ configuration.

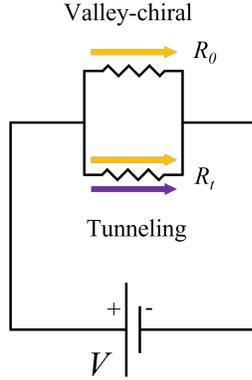

**Figure S2. Two-Channel Toy-Model of Electrical Transport in the "Non-trivial" Case.**

The top resistor represents the valley-chiral channel carrying current only from the K valley (yellow arrow); the bottom resistor represents trivial tunneling equally composed of the currents from both the K and K′ valleys (yellow and purple arrows).

### S3. Estimation of Valley Polarization at Zero $B$ Field

We use the parallel conducting channel toy-model described in S2 to obtain a rough estimation of the valley polarization of the total current. As elaborated in S2, we assume the ballistic conductance, $1/R_0$, to be valley-chiral and the tunnel conductance, $1/R_t$, to be non-chiral. With a bias voltage, $V$, applied, only carriers from the K valley are allowed across the chiral channel $1/R_0$ while carriers from both valleys contribute equally via non-chiral $1/R_t$. The total measured current therefore consists of $I_K = V/R_0 + V/(2R_t)$ and $I_{K'} = V/(2R_t)$. The valley polarization can then be estimated by $VP = I_K / (I_K + I_{K'}) = (1 + R/R_0)/2$. Beyond this rough estimation, a direct and accurate characterization of the valley polarization requires a network of four PJs with different valley chirality with a measurement scheme similar to previous experiments [4,5] on a gate-defined valley crossroad. Such a device architecture and measurement are future directions beyond the scope of this work.

### S4. Simulation of Tunneling Barriers and Illustration of Type-1 and Type-2 Tunnel Currents

To better understand the qualitative difference between the "trivial" (Fig. S1a) and "non-trivial" (Fig. S1b) configurations of a single PJ, a simulation of the lateral electrostatic configuration in the device is performed in COMSOL Multi-Physics. The geometric parameters in the simulation, such as the hBN thickness and gates' dimensions, match those of Device 1. The simulation is performed for a set of gate voltages matching some of those in Fig 2g. The horizontal axes in Fig. S1, a and b, indicate the lateral dimensions of the device near the PJ. In Fig. S1, we plot the spatial distribution of charge carrier density ($n$, indicated by the color) and the tunnel barrier height from the electrostatic simulation. The red and blue areas represent the P-doped and N-doped regions, respectively; the white color domains indicate the insulating regions. In the "trivial" regime, a finite tunnel barrier is present near the PJ, thus, Type-2 leakage current near the PJ can be suppressed with a larger $D$ field. In the "non-trivial" configuration, Type-2 leakage current near the PJ cannot be eliminated by larger $D$ fields as the bandgap always closes at the PJ. It can only be suppressed at a high magnetic field by establishing a tunnel barrier with the $v = 0$ Landau gap.

### S5. Results from Other Devices

Besides the devices (Device 1, Device 2) presented in the main manuscript, transport signatures of both "trivial" and valley-chiral PJs were observed in Device 3-7 (Fig. S3) with the resistances comparable to those in Device 1. Like Device 1, Device 3 was also measured under different displacement fields and DC biases (Fig. S4). Device 3 reproduced the behavior of Device 1: the resistance of the PJ in the "trivial" regime grew with the displacement field (Fig. S4a) and leveled off in the valley-chiral regime (Fig. S4b). Similarly, the non-linear (Fig. S4c) and Ohmic (Fig. S4d) dependence of the differential resistance on a DC bias was reproduced in Device 3 in the non-chiral and chiral regimes, respectively, for multiple displacement fields.

The resistance of the PJ in the "non-trivial" configuration in Device 5 (Fig. S5a) approaches the expected $h/(4e^2)$ under high magnetic fields, reproducing the main observation in magneto-transport result. Weak signatures of Shubnikov-de Haas oscillations are observed but are ill-defined compared to Device 1. We attribute this to realistic sample-dependent edge profiles. The top gates are used as part of the etching mask when defining the device boundary with reactive ion etching. Depending on the etch profile, a narrow region of the graphene may be gated ineffectively by the top gate. When conducting compressible strips (QH edge states) move toward the sample boundary at high magnetic fields, this narrow region may create a leakage path (Fig. S5b) around the $\nu = 0$ insulating regions and away from the PJ region.

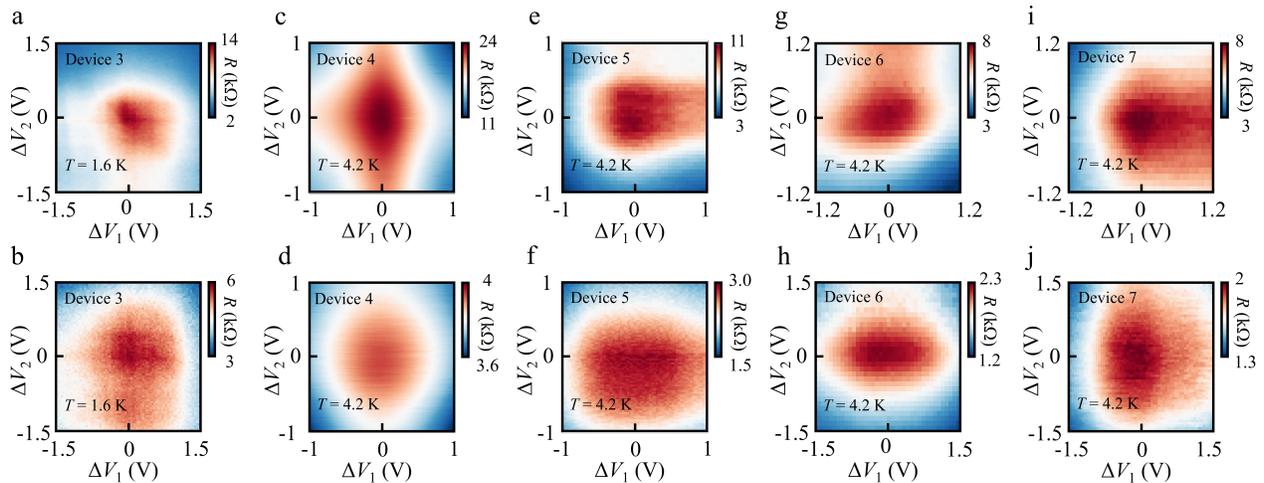

**Figure S3. Transport Signatures of a PJ in Multiple Devices.**
A PJ formed in additional samples (Device 3-7) in both "trivial" (a, c, e, g, i) and "non-trivial" (b, d, f, h, j) cases. In both scenarios, the resistance demonstrates a peak when a PJ is configured. The resistance in the "trivial" case is always significantly larger due to the absence of valley-chiral states.

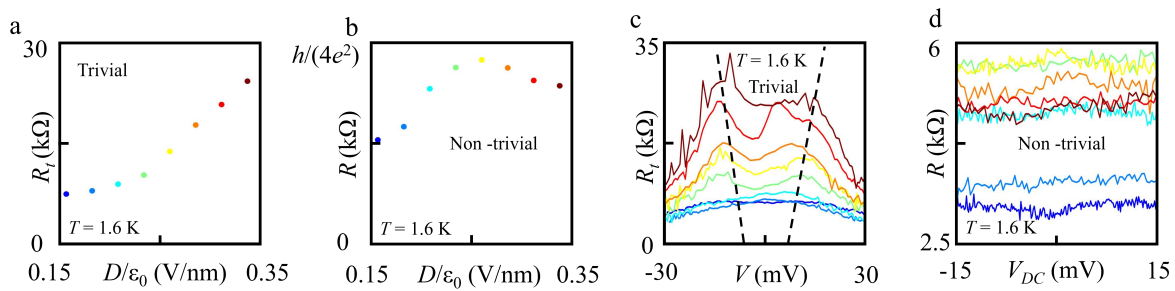

**Figure S4. Displacement Field Dependence of Resistance and DC Bias Measurements in Device 3.**
(a) PJ resistance as a function of displacement field (also indicated by the color) for the "trivial" configuration in Device 3. (b) PJ resistance as a function of displacement field for the "non-trivial" configuration in Device 3. (c) R-V curves at different displacement fields in Device 3 in the "trivial" configuration. The dashed lines show the position of the peaks and mark the size of half of the bandgap. (d) R-V curves at different displacement fields in Device 3 in the "non-trivial" configuration. In (c and d), $R_t$ and R represent the differential resistance and $V_{DC}$ is the DC bias voltage. The colors in (c and d) match those in (a and b).

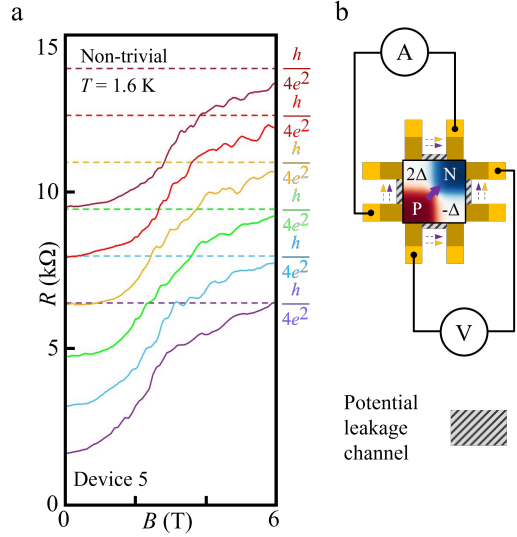

**Figure S5. Magneto-transport in Valley-Chiral PJ in Device 5.**
(a) Resistance of the valley-chiral configuration in Device 5 as a function of magnetic field at different $D$ fields ranging from 0.15 V/nm (purple) to 0.34 V/nm (dark red) with an increment of 0.04 V/nm. At high magnetic fields, the resistance approaches the predicted resistance, $h/(4e^2)$, of the valley-chiral states which is indicated by the dashed lines. The graphs are vertically offset by 1.5 kΩ for clarity. (b) Schematics of the device architecture and circuit diagram for Device 5 configured for four-probe measurements of the resistance. The insulating and doped regions forming a PJ are set up so that the valley-chiral current is carried by the K´ valley (purple arrow) in-gap states. The shaded area indicates the location of potential leakage paths along the etched edges depending on the etch profile that may vary from sample to sample.

## S6. Estimation of Carrier Mobility

By calculating the mean free path of the charge carriers in the P-doped region, we estimate the mobility of the charge carriers to be μ ~50000 cm$^2$/(V·s) by fitting the data with the Drude model $\sigma = |n|e\mu + \sigma_0$ [6-8], where $n$ is the carrier density, $e$ is the elementary charge, and $\sigma_0$ is the conductivity due to residual chemical doping of BLG. The corresponding mean free path [8] of the charge carriers is estimated by $l = \frac{h}{2e}\mu\sqrt{|n|/\pi}$ where $h$ is the Planck constant. In the bulk, ~100 nm away from the center PJ, $n$ is on the order of $10^{12}$ cm$^{-2}$ (with the main data taken close to $n = 3.75 \cdot 10^{12}$ cm$^{-2}$) and the mean-free path is estimated to be greater than 1 μm, an order of magnitude larger than the characteristic geometric size (~ 100 nm) of the point junction.